\documentclass{article}
\begin{document}

\title{Phase diagram of random lattice gases in the annealed limit}
\author{\textbf{A. P. Vieira}$^{\dagger}$\textbf{ and L. L. Gon\c{c}alves}$^{\ddagger}$\\
Departamento de F\'{\i}sica, Universidade Federal do Cear\'a\\
CP 6030, 60451-970 Fortaleza (CE) Brazil\\
$^{\dagger}$ {\small Present address: Instituto de F\'{\i}sica, Universidade de S\~ao Paulo}\\
{\small CP 66318, 05315-970 S\~{a}o Paulo (SP) Brazil }\\
$^{\dagger}$ {\small E-mail: apvieira@if.usp.br}\\
$^{\ddagger}$ {\small E-mail: lindberg@fisica.ufc.br}}
\maketitle
\begin{abstract}
An analysis of the random lattice gas in the annealed limit is presented.
The statistical mechanics of disordered lattice systems is briefly reviewed.
For the case of the lattice gas with an arbitrary uniform interaction
potential and random short-range interactions the annealed limit is
discussed in detail. By identifying and extracting an entropy of mixing
term, a correct physical expression for the pressure is explicitly given.
The one-dimensional lattice gas with uniform long-range interactions and
random short-range interactions satisfying a bimodal annealed probability
distribution is discussed. The model is exactly solved and is shown to
present interesting behavior in the presence of competition between
interactions, such as the presence of three phase transitions with different
critical temperatures and the occurrence of triple and quadruple points.
\end{abstract}

\label{intro}The lattice gas model \cite{lee52} was introduced by Lee and
Yang in 1952 as an application of their theory of condensation, due to its
equivalence to the Ising model. Despite its somewhat artificial character,
it has been extensively and successfully applied to the study of both
solid-liquid and liquid-gas transitions \cite{runnels72}. It has also been
applied to problems of adsorption of a gas on a crystal surface \cite
{oliveira78,ebner83,niskanen85}.

The mapping of the two-dimensional lattice gas with nearest-neighbor
interaction into the Ising model in the square lattice allowed the exact
calculation of the coexistence curve of a fluid transition exhibited by the
lattice gas at sufficiently low temperatures. However, the corresponding
equation of state could not be exactly determined, due to the absence of an
exact solution for the Ising model in the presence of an external magnetic
field. Such a solution exists for the one-dimensional analog of the model,
although it undergoes no phase transition at finite temperature, having very
little to contribute to even a qualitative analysis of the behavior of real
substances.

Notwithstanding, in the presence of uniform infinite-range interactions
phase transitions are induced even in one-dimensional systems \cite
{lebowitz66}. Furthermore, as pointed out by Hemmer and Stell \cite
{hemmer70,stell72}, for lattice and continuous fluids in which the potential
has both a long-range attraction and a short-range repulsion, plus a hard
core, up to two phase transitions are expected. This is indeed verified for
the $1d$ lattice gas with nearest-neighbor repulsion and long-range
attraction \cite{wilson77,vieira95}, described by the Hamiltonian 
\begin{equation}
H=-4J\sum_{j=1}^{N}n_{j}n_{j+1}-\frac{4I}{N}\sum_{i,j=1}^{N}n_{i}n_{j},
\label{unimod}
\end{equation}
where $J<0$, $I>0$ and the occupation numbers $n_{j}\in \{0,1\}$ represent
the absence or presence of a particle in the $j$th site along the chain.
Such model is mathematically equivalent to the Ising chain with short- and
long-range interactions in an external field \cite{nagle70}, which in turn is
equivalent to a linear chain approximation for higher dimensional models.
Depending on the ratio $\,$of the interaction parameters, the system can
present one or two phase transitions, and in the latter case a triple point
may be present \cite{wilson77,vieira95}. The three possible phases have
zero-temperature densities given by $\rho =0$ (gas phase), $\rho =\frac{1}{2}
$ (liquid phase) and $\rho =1$ (close-packed phase). However, the symmetry
between unoccupied and occupied cells (hole-particle symmetry) \cite
{rowlinson70} makes the critical temperatures for both transitions the same,
differently from what happens in the analogous continuous model \cite
{wilson77,kurioka88a}.

On the other hand, for random versions of the model, by looking at ground
state properties one can see that the presence of disorder in the
nearest-neighbor interactions induces both the breaking of the hole-particle
symmetry and the appearance of additional phase transitions at $T=0$. Due to
the presence of the long-range attraction those new transitions should
persist at low but finite temperatures. It is then worthwhile examining
disordered versions of the model and look for what new interesting
properties may arise.

The main purpose of this paper is to study the phase diagram of the $1d$
lattice gas with random annealed nearest-neighbor interactions and uniform
long-range interactions. This particular disordered version of the model \ref
{unimod} has the advantage of being exactly soluble. Although of difficult
physical realization, it might be useful in studying adsorption problems, if
one wishes to consider situations in which the presence of adsorbed
particles induces diffusion of crystalline atoms in or near imperfect
surfaces.

The outline of this paper is as follows. In order to clarify one subtle
aspect of the mathematical treatment of annealed models, a preceding
discussion is presented. Section \ref{disorder} discusses some theoretical
aspects of the statistical mechanics of disordered systems. The $d$%
-dimensional lattice gas with an arbitrary uniform interaction potential
plus random short-range interactions is treated in Sec. \ref{lgrsrint},
following the approach of Thorpe and Beeman for the Ising model \cite
{thorpe76}. The calculation of thermodynamic functions for this class of
systems is presented, and a general method for calculating the equation of
state is proposed. Then the one-dimensional lattice gas with random
nearest-neigbor interactions and uniform long-range interactions is
considered in Sec. \ref{slrint}. The main features of the behavior of the
model are presented and discussed when the nearest-neighbor interactions are
selected from a bimodal distribution. In the final section the conclusions
and main results of the paper are summarized. The equivalence between the
model and the random Ising model is presented in Appendix A, and in Appendix
B the ground state properties of the model are discussed.

\section{Disordered systems.}

Disordered systems are characterized by two kinds of variables: dynamical
variables, with mean relaxation time $\tau _{h}$, and ``structural''
variables, related to the randomness of the system and with mean relaxation
time $\tau _{c}$. As introduced by Brout \cite{brout59}, there are two limits
in which disorder problems can be formally treated, namely, the annealed and
the quenched limits. The latter is appropriate to systems where disorder
relaxes in a much longer time scale than dynamical variables ($\tau _{c}\gg
\tau _{h}$), and so disorder variables can be considered as effectively
frozen. The opposite situation, in which disorder and dynamical variables
fluctuate in the same time scale ($\tau _{c}\approx \tau _{h}$), corresponds
to the annealed limit. Since exact solutions can be found for some models,
in these cases it is worthwhile using the annealed limit as an approximation
to a quenched system, as it will be further discussed.

\label{disorder}As shown by Mazo \cite{mazo63}, the free energy of a random
quenched system is given by 
\begin{equation}
F_{q}=-k_{B}T\sum_{\left\{ \kappa \right\} }p(\kappa )\ln Q(\kappa
)+k_{B}T\sum_{\left\{ \kappa \right\} }p(\kappa )\ln p(\kappa ),  \label{fq}
\end{equation}
where $k_{B}$ is Boltzmann's constant, $T$ is the absolute temperature of
the system, $p(\kappa )$ is the fixed probability of occurrence of the
random configuration $\{\kappa \}$, $Q(\kappa )$ is the partition function
of the system for that particular configuration, and the summations run over
all possible configurations $\{\kappa \}$. The second summation in Eq.\ref
{fq} is interpreted as due to an entropy of mixing of the random variables.
It is in general ignored, since the only way to measure an entropy is to
change it, which is not possible in this case because the probabilities $%
p(\kappa )$ are rigid by assumption. Irrespective of that term, in general
one finds that, for almost all thermodynamic functions obtained from the
free energy by differentiation, the average value $X_{q}$ of the function
can also be calculated as an average of the corresponding values $X(\kappa )$
over all the disorder configurations, 
\begin{equation}
X_{q}=\sum_{\left\{ \kappa \right\} }p(\kappa )X(\kappa ).
\end{equation}

When lattice models are considered, disorder is usually represented by site
or bond random variables. For example, in problems of adsorption of gas
molecules on a crystal surface one may want to consider sites with random
potentials or random interactions between particles in different adsorption
sites. For simplicity, consider a lattice of $N$ sites (or bonds). Suppose
that some random variable $\kappa _{j}$ can take one of $M$ values $%
\{K_{1},K_{2},\ldots ,K_{M}\}$ at a given site (or bond) $j$ with
probability $q_{i}$, such that $\sum_{i=1}^{M}q_{i}=1$. Since that
probability is independent of the values of the other random variables in
the system, the probability of a particular configuration $\{\kappa \}\equiv
\{\kappa _{1},\kappa _{2},\ldots ,\kappa _{N}\}$ is given by 
\begin{equation}
p(\kappa )=\prod_{i=1}^{M}q_{i}^{n_{i}},
\end{equation}
where $n_{i}$ is the number of sites (or bonds) to which the value $K_{i}$
of the random variable is associated in that given configuration $\{\kappa
\} $. Obviously the $n_{i}$'s obey the constraint 
\begin{equation}
\sum_{i=1}^{M}n_{i}=N.
\end{equation}
It is possible to show that for this case the entropy of mixing in Eq.\ref
{fq} takes the form 
\begin{equation}
-k_{B}\sum_{\left\{ \kappa \right\} }p(\kappa )\ln p(\kappa
)=-Nk_{B}\sum_{i=1}^{M}q_{i}\ln q_{i},
\end{equation}
which is proportional to the volume of the system. So, if one keeps the
entropy of mixing term and tries to calculate (in appropriate units) the
pressure of a quenched lattice gas the result is 
\begin{equation}
P_{q}=-\frac{\partial F_{q}}{\partial N}=k_{B}T\left[ \sum_{\left\{ \kappa
\right\} }p(\kappa )P(\kappa )-\sum_{i=1}^{M}q_{i}\ln q_{i}\right] ,
\label{pqe}
\end{equation}
where $P(\kappa )$ is the pressure calculated for a particular configuration 
$\left\{ \kappa \right\} $. Clearly Eq.\ref{pqe} is a physically incorrect
expression for the pressure, since in the zero density limit the first term
vanishes, while the second term is positive. Furthermore, in the non-random
limit ($K_{i}\equiv K$ for all $i$) all configurations are identical, but
the entropy of mixing still contributes to the ``pressure''. Finally the
entropy of mixing diverges in the continuum limit ($M\rightarrow \infty $, $%
q_{i}\rightarrow 0$)  \cite{thorpe76}. Thus the correct expression for the
pressure must lack the entropy of mixing, taking the form 
\begin{equation}
P=k_{B}T\sum_{\left\{ \kappa \right\} }p(\kappa )P(\kappa ),
\end{equation}
which is the expression used in previous studies of random lattice gases \cite
{inawashiro81b,ausloos83}. The need to drop out the entropy of mixing in the
calculation of the pressure of a quenched fluid was also pointed out by
Singh and Kovac \cite{singh89} on the basis of the interpretation of that
term as an information entropy, introduced by Sobotta and Wagner \cite
{sobotta79}.

For most systems the quenched limit is more realistic than the annealed
limit, since in general $\tau _{c}\gg \tau _{h}$, as it happens in spin
glasses, for instance. However, the exact mathematical treatment of the
quenched limit is very difficult, due to the fact that the calculation of
the free energy involves averaging logarithms of partition functions. On the
other hand, the annealed free energy can be written in the form 
\begin{equation}
F_{a}=-k_{B}T\ln Q_{a}  \label{fa}
\end{equation}
where $Q_{a}$ denotes the average of the partition function over the
disorder configurations. This is often calculated with the constraint that
the thermal averages of the random variables obey some prescribed
distribution, which is realized by the introduction of Lagrange multipliers
playing the role of pseudo-chemical potentials. This implies that in the
annealed limit the disorder variables are adjusted so as to minimize the
free energy, a procedure which singles out a subset of the disorder
configurations and introduces correlations between the disorder variables
themselves. So, the annealed limit is in general not a good approximation to
the quenched limit. Nevertheless, Morita \cite{morita64} showed that a
quenched system can be represented by a fictitious (annealed) equilibrium
system subject to an additional potential, composed of an infinite series of
Lagrange multipliers adjusted by the (quenched) $n$-``particle'' correlation
functions ($n=1,2,3,\ldots $) for the random variables. Indeed, for various
Ising models it can be verified that by controlling correlations between
pairs of random variables ($n=2$) results for annealed systems approach
those of the corresponding quenched systems \cite{thorpe78,serva93}. This
treatment introduces mathematical difficulties which are of course
increasingly greater, and so this work will analyze the usual case in which
one tries to control only the concentration of the random variables ($n=1$).

In the next section it will be shown that an entropy of mixing term can also
be identified in a large class of annealed models, and must likewise be
discarded in order to produce meaningful results.

\section{Lattice gas with random short-range interactions: annealed limit.}

\label{lgrsrint}Consider a $N$-site $d$-dimensional lattice gas model whose
particles interact via an arbitrary uniform (non-random) potential plus
short-range interactions satisfying the annealed probability distribution 
\begin{equation}
\wp (\kappa _{ij})=\sum_{k=1}^{M}q_{k}\delta (\kappa _{ij}-J_{k}),
\end{equation}
for all pairs of nearest-neighbor sites $i$ and $j$. The Hamiltonian of the
model can be written as 
\begin{equation}
H=H^{(U)}+\sum_{<i,j>}H_{ij},
\end{equation}
where $H^{(U)}$ is the uniform term, the summation runs over all
nearest-neighbor pairs of sites and 
\begin{equation}
H_{ij}=-\sum_{k=1}^{M}\kappa _{ij}n_{i}n_{j}.
\end{equation}
For any site $i$, variable $n_{i}=0$ if the site is empty or $n_{i}=1$ if
the site is occupied by a particle.

Following the approach of Thorpe and Beeman for the Ising model  \cite
{thorpe76}, this situation can be formally treated by writing $\kappa _{ij}$
in the form 
\begin{equation}
\kappa _{ij}=\sum_{k=1}^{M}t_{ij,k}J_{k},
\end{equation}
and introducing $M$ pseudo-chemical potentials $\xi _{k}$ to control the
averages of the random numbers $t_{ij,k}$, which for all $(i,j)$ are subject
to the constraints 
\begin{equation}
t_{ij,k}\in \left\{ 0,1\right\} ,\mbox{\quad }\sum_{k=1}^{M}t_{ij,k}=1,\mbox{%
\quad and\quad }\left\langle t_{ij,k}\right\rangle =q_{k}.  \label{vinc}
\end{equation}

The annealed grand partition function of the system is given by 
\begin{equation}
\mathcal{Z}=\sum_{\{t\}}{}^{^{\prime }}\exp \left[ \beta \sum_{k=1}^{M}\xi
_{k}\sum_{<i,j>}t_{ij,k}\right] \sum_{\{n\}}\exp \left[ \beta \mu
\sum_{j=1}^{N}n_{j}-\beta H\right] ,
\end{equation}
where the primed summation runs over those configurations satisfying the
first two constraints in \ref{vinc}. After performing the partial trace over
the random variables one obtains 
\begin{equation}
\mathcal{Z}=\sum_{\{n\}}\exp \left[ \beta \mu \sum_{j=1}^{N}n_{j}-\beta
H^{(U)}\right] \prod_{<i,j>}\left\{ \sum_{k=1}^{M}\exp \left[ \beta \left(
\xi _{k}+J_{k}n_{i}n_{j}\right) \right] \right\} .  \label{zanp}
\end{equation}
Due to the fact that $n_{i}\in \left\{ 0,1\right\} $ for all sites the term
in braces in Eq.\ref{zanp} can be expressed as 
\begin{equation}
\sum_{k=1}^{M}\exp \left[ \beta \left( \xi _{k}+J_{k}n_{i}n_{j}\right) %
\right] \equiv Ae^{Kn_{i}n_{j}},
\end{equation}
if $A$ and $K$ are defined by the expressions 
\begin{equation}
A=\sum_{k=1}^{M}w_{k}\quad \mbox{and\quad }e^{K}=\left.
\sum_{k=1}^{M}w_{k}e^{\beta J_{k}}\right/ \sum_{k=1}^{M}w_{k}.  \label{ak}
\end{equation}
with $w_{k}\equiv \exp (\beta \xi _{k})$. Equation \ref{zanp} can now be
written as 
\begin{equation}
\mathcal{Z}=A^{Nq/2}\sum_{\{n\}}\exp \left[ \beta \mu
\sum_{j=1}^{N}n_{j}-\beta H^{(U)}+\sum_{<i,j>}Kn_{i}n_{j}\right] \equiv
A^{Nq/2}\mathcal{Q}(K),  \label{zak}
\end{equation}
where $q$ is the coordination number of the lattice and $\mathcal{Q}(K)$ is
just the grand partition function of a regular lattice gas with short-range
interaction energy $K/\beta $ plus the uniform interaction potential.

The fugacities $w_{k}$ are eliminated by imposing for all nearest-neighbor
pairs $(i,j)$ and all values of $k$ the condition 
\begin{equation}
\left\langle t_{ij,k}\right\rangle =q_{k}\Rightarrow \frac{w_{k}}{N}\frac{%
\partial \ln \mathcal{Z}}{\partial w_{k}}=q_{k},
\end{equation}
which can be rewritten as 
\begin{equation}
w_{k}\frac{\partial \ln A}{\partial w_{k}}+\epsilon (K)w_{k}\frac{\partial K%
}{\partial w_{k}}=q_{k},  \label{qak}
\end{equation}
where 
\begin{equation}
\epsilon (K)\equiv \left\langle n_{i}n_{j}\right\rangle _{K}=\frac{2}{Nq}%
\frac{\partial \ln \mathcal{Q}(K)}{\partial K}
\end{equation}
is the nearest neighbor pair correlation function for the regular lattice
gas with interaction $K/\beta $. By rearranging terms in Eq.\ref{qak} and
imposing the condition $\sum_{k=1}^{M}q_{k}=1$ one obtains 
\begin{equation}
\sum_{k=1}^{M}\frac{q_{k}}{\coth \left[ \frac{1}{2}(K-\beta J_{k})\right]
+1-2\epsilon (K)}=0,  \label{kek}
\end{equation}
which determines $K$ if one is able to calculate $\epsilon (K)$. Notice that
in general the effective interaction $K/\beta $ depends on the temperature.

The density of the gas is set by the condition 
\begin{equation}
\rho =\frac{z}{N}\frac{\partial \ln \mathcal{Z}}{\partial z}=\frac{z}{N}%
\frac{\partial \ln \mathcal{Q}(K)}{\partial z},
\end{equation}
and from this equation it is possible to eliminate the fugacity $z$.

The annealed free energy is given by  \cite{huang87} 
\begin{equation}
f_{a}=-k_{B}T\left[ \frac{1}{N}\ln \mathcal{Z}-\sum_{k=1}^{M}q_{k}\ln
w_{k}-\rho \ln z\right] ,
\end{equation}
which after straightforward calculations can be written in the form 
\begin{equation}
f_{a}=f_{K}-k_{B}T\left\{ \sum_{k=1}^{M}q_{k}\ln \left[ 1-\epsilon (K)\left(
1-e^{\beta J_{k}-K}\right) \right] -\sum_{k=1}^{M}q_{k}\ln q_{k}\right\} ,
\label{fanri}
\end{equation}
where 
\begin{equation}
f_{K}=-k_{B}T\left[ \frac{1}{N}\ln \mathcal{Q}(K)-\rho \ln z\right]
\end{equation}
is the free energy of the regular system. The last summation on the right
hand side of Eq.\ref{fanri} is an entropy of mixing term, identical to that
in Eq.\ref{pqe}, and must be dropped as discussed in Sec.\ref{disorder}.
With that correction, the free energy per lattice site takes the form 
\begin{equation}
f=f_{K}-k_{B}T\sum_{k=1}^{M}q_{k}\ln \left[ 1-\epsilon (K)\left( 1-e^{\beta
J_{k}-K}\right) \right] ,  \label{fnoei}
\end{equation}
and the pressure of the system can then be calculated from the thermodynamic
relation 
\begin{equation}
P=\rho \mu -f=k_{B}T\rho \ln z-f\Rightarrow
\end{equation}
\begin{equation}
\Rightarrow P=k_{B}T\left\{ \frac{1}{N}\ln \mathcal{Q}(K)+%
\sum_{k=1}^{M}q_{k}\ln \left[ 1-\epsilon (K)\left( 1-e^{\beta
J_{k}-K}\right) \right] \right\} ,  \label{pri}
\end{equation}
the first term being just the pressure of the regular lattice gas with
effective interaction $K/\beta $. In the non-random limit $(J_{k}\equiv J$
for all $k)$ Eq.\ref{kek} gives $K=\beta J$ and according to Eq.\ref{pri}
the pressure is given by 
\begin{equation}
P\equiv P_{K}=\frac{k_{B}T}{N}\ln \mathcal{Q}(K),
\end{equation}
which is obviously the correct expression.

From the results of this section it is then clear that, as in the quenched
limit, the entropy of mixing term can be made explicit in the free energy of
any lattice gas with annealed short-range interactions, and that the correct
equation of state can be obtained by discarding that term. This is a fairly
general result and will be applied in the next section to the $1d$ lattice
gas with random nearest-neighbor and uniform long-range interactions.

\section{One-dimensional lattice gas with random short-range interactions
and uniform long-range interactions: annealed limit.}

\label{slrint}The existence of exact solutions \cite{wilson77, vieira95} for
the uniform model \ref{unimod} allows one to exactly solve the model in the
presence of random annealed nearest-neighbor interactions, as shown in the
previous section.

The Hamiltonian of a $1d$ lattice gas with uniform long-range interactions
and random nearest-neighbor interactions is given by 
\begin{equation}
H=-4\sum_{j=1}^{N}\kappa _{j}n_{j}n_{j+1}-\frac{4I}{N}%
\sum_{i,j}^{N}n_{i}n_{j},  \label{hran}
\end{equation}
where the nearest-neighbor interactions $\kappa _{j}\,$satisfy the annealed
distribution 
\begin{equation}
\wp (\kappa _{j})=\sum_{k=1}^{M}q_{k}\delta (\kappa _{j}-J_{k}).
\end{equation}

The thermodynamic properties of this random model can be obtained from those
of a reference system without the long-range interactions. In particular the
Helmholtz free energy is given by \cite{lebowitz66} 
\begin{equation}
f(\rho ,T)=\mbox{CE}\left[ \tilde{f}(\rho ,T)-4I\rho ^{2}\right] ,
\end{equation}
where $\tilde{f}(\rho ,T)$ is the free energy of the reference system as a
function of the density $\rho $ and temperature $T$, while ``CE'' denotes
the convex envelope of the function in square brackets. Correspondingly, the
pressure $P$ and chemical potential $\mu $ are given by \cite{lebowitz66} 
\begin{equation}
P=\mbox{MC}\left[ \tilde{P}-4I\rho ^{2}\right] ,\quad \mu =\tilde{\mu}%
-8I\rho ,  \label{plr}
\end{equation}
where ``MC'' denotes the Maxwell construct of the function in square
brackets and 
\begin{equation}
\tilde{P}=\rho \tilde{\mu}-\tilde{f}(\rho ,T)\quad \mbox{and}\quad \tilde{\mu%
}=\frac{\partial \tilde{f}}{\partial \rho }
\end{equation}
are the corresponding functions for the reference system.

From the results of the previous section, Eqs.\ref{fnoei} and \ref{pri}, the
free energy $\tilde{f}$ and pressure $\tilde{P}$ of the reference system
without long-range interaction are given in terms of the corresponding
functions $\tilde{f}_{K}$ and $\tilde{P}_{K}$ for a regular model with
effective nearest-neighbor interaction $K/\beta $. The grand-partition
function of this regular model is given by 
\begin{equation}
\tilde{\mathcal{Q}}(K)=\sum_{\{n\}}\exp \left[ \beta \tilde{\mu}%
\sum_{j=1}^{N}n_{j}+K\sum_{j=1}^{N}n_{j}n_{j+1}\right] ,
\end{equation}
and by using the transfer matrix method with cyclic boundary conditions one
easily finds 
\begin{equation}
\beta \tilde{P}_{K}=\frac{1}{N}\ln \tilde{\mathcal{Q}}(K)=\ln \frac{1}{2}%
\left\{ 1+\tilde{z}e^{4K}+\left[ \left( 1-\tilde{z}e^{4K}\right) ^{2}+4%
\tilde{z}\right] ^{1/2}\right\} ,  \label{pref}
\end{equation}
where $\tilde{z}\equiv \exp \left( \beta \tilde{\mu}\right) $. The density
of the system is then given by the relation 
\begin{equation}
\rho =\frac{\partial \tilde{P}_{K}}{\partial \tilde{\mu}}=\tilde{z}\left\{ 
\frac{e^{4K}-\left[ \left( 1-\tilde{z}e^{4K}\right) e^{4K}-2\right] \left[
\left( 1-\tilde{z}e^{4K}\right) ^{2}+4\tilde{z}\right] ^{-1/2}}{1+\tilde{z}%
e^{4K}+\left[ \left( 1-\tilde{z}e^{4K}\right) ^{2}+4\tilde{z}\right] ^{1/2}}%
\right\} ,  \label{dens}
\end{equation}
from where the fugacity $\tilde{z}$ of the reference system can be
determined, and the nearest-neighbor pair correlation function of the
regular system is given by 
\begin{equation}
\epsilon (K)=\frac{\partial \left( \beta \tilde{P}_{K}\right) }{\partial K}=%
\tilde{z}e^{4K}\left\{ \frac{1-\left( 1-\tilde{z}e^{4K}\right) \left[ \left(
1-\tilde{z}e^{4K}\right) ^{2}+4\tilde{z}\right] ^{-1/2}}{1+\tilde{z}e^{4K}+%
\left[ \left( 1-\tilde{z}e^{4K}\right) ^{2}+4\tilde{z}\right] ^{1/2}}%
\right\} .  \label{correl}
\end{equation}

Finally, the pressure of the random system with long-range interactions
described by the Hamiltonian in Eq.\ref{hran} is 
\begin{equation}
P=\mbox{MC}\left\{ \tilde{P}_{K}+k_{B}T\sum_{k=1}^{M}q_{k}\ln \left[
1-\epsilon (K)\left( 1-e^{4\beta J_{k}-4K}\right) \right] -4I\rho
^{2}\right\} ,  \label{pran}
\end{equation}
with $K$ determined from the solution of Eq.\ref{kek}. So, the complete
equation of state can be obtained from Eqs.\ref{pref}-\ref{pran}.

An alternative approach can be used to solve this problem, namely, the
mapping onto the random Ising model. This is discussed in Appendix \ref
{equivalence} and has the advantage of making evident the breaking of
hole-particle symmetry. Furthermore, it provides a way of avoiding the use
of the Maxwell construct by previously locating free energy minima, which
directly indicate phase transitions.

The results above are applied to the case in which the nearest-neighbor
interactions are selected from the bimodal distribution 
\begin{equation}
\wp (\kappa _{j})=p\delta (\kappa _{j}-J_{A})+(1-p)\delta (\kappa
_{j}-J_{B}),
\end{equation}
where the interactions $J_{B}$ are assumed repulsive ($J_{B}<0$). This is
done in order to introduce competition effects, which, as in the pure model,
are responsible for complex behavior. In the absence of competition between
short- and long-range interactions the system presents at most one phase
transition. The ground state properties of the model, discussed in detail in
Appendix \ref{ground}, indicate that at $T=0$ and for $J_{A}\neq 0$ the
system undergoes at most three phase transitions. For $J_{A}>0$, when there
is competition between the nearest-neighbor interactions, at fixed pressure
the system can present itself in structures characterized by $\rho =0$ (gas
phase), $\rho =p$, $\rho =\frac{1}{2}(1+p)$ (liquid phases), and $\rho =1$
(close-packed phase). For $J_{A}<0$ all nearest-neighbor interactions are
repulsive, and the stable phases are those for which $\rho =0$ (gas), $\rho =%
\frac{1}{2}$, $\rho =\frac{1}{2}(1+p)$ (liquid phases) and $\rho =1$
(close-packed). For $J_{A}=0$ the system undergoes at most two transitions,
associated with those of the pure model  \cite{wilson77,vieira95}, the
liquid-phase density being shifted to $\rho =\frac{1}{2}(1+p)$.

The characterization of the phases $\rho =\frac{1}{2}$, $\rho =p$ and $\rho =%
\frac{1}{2}(1+p)$ as liquid phases is based on the instability of the
corresponding structures. As in the uniform case, the $\rho =\frac{1}{2}$
phase at $T=0$ is stable at certain pressures because particles are forced
by nearest-neighbor repulsion to occupy every other site of the lattice. At
any finite temperature, thermal fluctuations overcome the effect of the
repulsion, inducing local fluctuations of the density and destroying the
structure. The stability of the $\rho =p$ and $\rho =\frac{1}{2}(1+p)$
phases at $T=0$, on the other hand, is related to the existence of
interaction domains (Appendix \ref{ground}). These are a consequence of the
interactions mobility in the annealed limit and are formed as particles are
added to the system at fixed volume in order to minimize the free energy. In
the $\rho =p$ phase all sites in the $J_{A}$ domains are occupied, while all
those in $J_{B}$ domains are empty. For $\rho =\frac{1}{2}(1+p)$ all sites
in $J_{A}$ domains and every other site in $J_{B}$ domains are occupied.
However, interaction domains are also destroyed by thermal flucuations at
any finite temperature, making the corresponding structures unstable.

In the following subsections the properties of the model at all temperatures
are discussed for the various possible cases. The relevant parameters are
renormalized by $\left| J_{B}\right| $, 
\begin{equation}
\delta \equiv \frac{J_{A}}{\left| J_{B}\right| },\quad \alpha \equiv \frac{I%
}{\left| J_{B}\right| },\quad \theta \equiv \frac{k_{B}T}{\left|
J_{B}\right| },\quad P^{\ast }=\frac{P}{\left| J_{B}\right| }.
\end{equation}

In passing it should be mentioned that, for repulsive long-range interaction
($I<0$), the system does not undergo a phase transition irrespective of the
sign of the short-range interaction.

\subsection{Case $J_{A}=0$.}

For dilute short-range interactions there are at most two phase transitions,
whose critical temperatures are different for $p\in (0,1)$ because of the
breakdown of hole-particle symmetry, as already stated. The critical
temperature $T_{c}$ of the gas-liquid (G-L) transition is always higher than 
$T_{c}^{\prime }$, the critical temperature of the liquid-close-packed
(L-CP) transition. This is in agreement to what is verified for the
corresponding non-random continuum model \cite{wilson77,kurioka88a}, in which
there is no hole-particle symmetry.

For $p\ne 1$ there is always a range of values of $\alpha $ in which one or
two triple points exist, as can be seen from the reentrant behavior of the
triple point temperature ($T_{t}$) curves in the $T\times \alpha $ diagrams
of Figs. 1(a)-(c), for $p=0.1$, $p=0.25$ and $p=0.5$, respectively. The
critical temperature of the G-L transition coincides with the triple point
temperature for $\alpha =\alpha _{ct}$, whose dependence on $p\,$is shown in
Fig. 1(d). It can be seen that $\alpha _{ct}$ approaches $\alpha _{tr}\simeq
3.1532$ as $p\rightarrow 0$, while for $p\rightarrow 1$, where there is no
short-range repulsion, $\alpha _{ct}$ approaches $\alpha _{0}=2$. This is
the value of $\alpha $ for which $T_{t}$ vanishes for all $p$ when $J_{A}=0$.

Examples of the behavior of the isotherms and phase diagrams for $p=0.1$ are
shown in Figs. 2-4. In Fig. 2, for $\alpha =1.5$, there are two-phase
transitions even at $T=0$ and there is no triple point. For low temperatures
the derivative of the L-CP transition pressure with respect to $T$ is
negative. This is related to the destruction by thermal fluctuations of the
interaction domains built at $T=0$, which makes the effective short-range
repulsion weaker, lowering the pressure needed to pack the particles. In
Fig. 3, for $\alpha =2.2$, there is only one phase transition (G-CP) at $T=0$%
, although two transitions occur for a range of temperatures above the
triple point. In Fig. 4, for $\alpha =\alpha _{ct}\simeq 2.4297$, the
critical temperature of the L-CP transition coincides with the triple point
temperature, and for $\alpha >\alpha _{ct}$ there is only one phase
transition. The behavior of the system when there are two triple points will
be exemplified in the next subsection.

\subsection{Case $J_{A}>0$.}

When there is competition between short-range interactions even a small
amount of randomness is responsible for interesting behavior, as can be seen
in the $\rho \times T$ projection of the coexistence surfaces shown in Figs.
5(a) and (b) for $J_{A}=-J_{B}$ ($\delta =1$), $p=0.07$ and two slightly
different values of $\alpha $. In both cases there are three phase
transitions at $T=0$, defining in decreasing density the close-packed, the
two liquid (L$_{1}$ and L$_{2}$) and the gas phases. The lowest critical
temperature corresponds to a L$_{2}$-G transition. Above that there is an
intermediate range of temperatures in which the system presents only two
transitions. For $\alpha =1.13$ that range is limited by a triple point,
immediately above whose temperature there is again three phase transitions,
as shown in Fig. 5(a). But for $\alpha =1.129$ that triple point is replaced
by another critical point, giving rise to a coexistence ``bubble'', as can
be seen in Figs. 5(b). The $T\times \alpha $ diagram for $\delta =1$ and $%
p=0.07$ is shown in Fig. 6. When the long-range attraction is sufficiently
weak the system can present three phase transitions, whose critical
temperatures are shown as solid lines in the figure. Triple points (broken
lines) are present for intermediate values of $\alpha $. The breaking and
the reentrant behavior of the $T_{c}$ curve, evident in the inset of Fig. 6,
is related to the appearance of coexistence ``bubbles''. That unusual
behavior is observed only for values of $p$ between $p\simeq 0.06$ and $%
p\simeq 0.08$.

The system may also present a quadruple point, i.e. coexistence of four
phases, if $J_{A}>-J_{B}$ ($\delta >1$). This is exemplified for $\delta =2$
and $p=0.1$ in the $T\times \alpha $ diagram in Fig. 7, where four triple
point curves (dotted lines) intersect at the quadruple point $P_{q}$.
Figures 8 and 9 show $P\times T$ and $\rho \times T$ diagrams for $\delta =2$%
, $p=0.08$ and values of $\alpha $ around $\alpha _{q}$, illustrating the
possible existence of two triple points or a quadruple point.

\subsection{Case $J_{B}<J_{A}<0\mathbf{.}$}

As detailed in Appendix B, when all short-range interactions are repulsive
the possible structures at $T=0$ correspond to $\rho =1$ (close-packed
phase), $\rho =\frac{1}{2}(1+p)$ (liquid-1 phase), $\rho =\frac{1}{2}$
(liquid-2 phase) and $\rho =0$ (gas phase).

When $\left| J_{B}\right| $ is much larger than $\left| J_{A}\right| $ the
behavior of the system is expected to be similar to the dilute case ($%
J_{A}=0 $). This is confirmed for the case $J_{A}=\frac{3}{10}J_{B}$ ($%
\delta =-\frac{3}{10}$), as shown in the $\theta \times \alpha $ diagrams in
Figs. 10(a)-(c) and the $\alpha \times p$ diagram in Fig. 10(d). Except for
the presence of three phase transitions for small values of $\alpha $, the
behavior is similar to that observed in Fig. 1. Isotherms and phase diagrams
are shown in Figs. 11 and 12 for $p=0.5$, $\alpha =0.8$ and $p=0.75$, $%
\alpha =0.8$, respectively. As expected for $\delta >-\frac{1}{2}\,$(see
Appendix B) the two-transition regime at $T=0$ consist of the $\rho =0$, $%
\rho =\frac{1}{2}(1+p)$ and $\rho =1$ phases, as in the dilute case. For
small values of the long-range interactions, the great energy difference
between short-range interactions leads to the high pressure needed to
close-pack the system, when it is necessary to occupy nearest neighbor sites
with $J_{B}$ bonds. At low pressure the occupation of those sites is
virtually forbidden, giving rise to the behavior shown in Figs. 11 and 12.
As the value of $\alpha $ is increased, the system becomes more similar to
the dilute case, as can be seen in Figs. 10(a)-(c).

When $\frac{1}{2}\left| J_{B}\right| <\left| J_{A}\right| <\frac{2}{3}\left|
J_{B}\right| $ ($-\frac{2}{3}<\delta <-\frac{1}{2}$) both two-transition
regimes at $T=0$, described in Appendix B are possible, with the
liquid-phase corresponding to $\rho =\frac{1}{2}$ or $\rho =\frac{1}{2}(1+p)$
if $p$ is less or greater than $p^{\prime }\equiv (1+2\delta )/(1+\delta )$,
respectively. At finite temperatures the case $p=p^{\prime }$ contains
ingredients of both regimes. The $\theta \times \alpha $ diagram for the
case $J_{A}=\frac{3}{5}J_{B}$ ($\delta =-\frac{3}{5}$) and $p=p^{\prime }=%
\frac{1}{2}$ is shown in Fig. 13. Notice that both triple point lines touch
the $\alpha $ axis at the same point, indicating the absence of a
two-transition regime at $T=0$. Figure 14 shows isotherms and phase diagrams
for the same case and $\alpha =1.45$, where a triple point is observed,
corresponding to the coexistence of the close-packed and the two liquid
phases. The L$_{1}$-L$_{2}$ transition line has a positive temperature
derivative, as shown in Fig. 14(b), differently from what is observed for $%
\delta >0$, as shown in Fig. 8(a).

For $p<p^{\prime }$ and in the two-transition regime, the $T=0$
liquid-close-packed transition pressure is given by $%
P=-4[pJ_{A}+(1-p)J_{B}+I]$, the arithmetic mean of the corresponding results
for the pure cases $p=0$ and $p=1$. For $p>p^{\prime }$ the system can
present a quadruple point, as observed for the $\delta >1$ case.

\section{Conclusions.}

The Helmholtz free energy of any lattice gas model with random short-range
interactions in the annealed limit was shown to present an entropy of mixing
term. As shown for the quenched limit, that term has to be dropped from the
free energy in order to give correct physical expressions for the pressure.

As an application, the one-dimensional lattice gas with uniform infinite
range interactions and random short-range interactions satisfying bimodal
distributions was exactly solved. Explicit results were presented for the
cases $J_{A}=0$, $J_{A}=-J_{B}>0$, $J_{A}=-2J_{B}>0$, $J_{A}=\frac{3}{10}%
J_{B}<0$ and $J_{A}=\frac{3}{5}J_{B}<0$, where there is competition between
short- and long-range interactions. In general, the system can present at
most three phase transitions at a fixed temperature. The phases were
identified as close-packed, gas and two liquid phases, the latter being
characterized at low temperatures by the existence of interaction domains. A
very important effect of disorder was the breaking of the hole-particle
symmetry of the pure model, which induces different critical temperatures
for the transitions. Under certain conditions phase diagrams present
negative thermal derivatives of some transition pressures, a phenomenon
observed in nature in water, for instance. The existence of triple and
quadruple points was demonstrated, and for $J_{A}>0$ and small values of $p$
the existence of coexistence ``bubbles'' was observed.
\medskip
\begin{center}
{\Large Appendix A. Equivalence between the random lattice gas and the random Ising model.}
{\large \medskip}
\end{center}
\label{equivalence}Here an alternative treatment for the problem of the
random lattice gas discussed in Sec. \ref{slrint} is presented.

By using the transformation \cite{lee52} 
\begin{equation}
n_{j}=(1-\sigma _{j})/2\mbox{,\quad }\sigma _{j}=\pm 1,  \label{lgim}
\end{equation}
it can be shown that the annealed grand partition function of the random
lattice gas, 
\begin{equation}
\mathcal{Z}=\sum_{\{t\}}{}^{^{\prime }}\exp \left[ \beta \sum_{k=1}^{M}\xi
_{k}\sum_{j=1}^{N}t_{j,k}\right] \sum_{\{n\}}\exp \left[ \beta \mu
\sum_{j=1}^{N}n_{j}-\beta H\right] ,
\end{equation}
can be written as 
\begin{equation}
\mathcal{Z}=\exp \left[ -N\left( \bar{h}+2\sum_{k=1}^{M}q_{k}J_{k}+\bar{I}%
\right) \right] Z_{\sigma },
\end{equation}
where $\bar{I}\equiv \beta I$, $\bar{h}\equiv \beta h=-2\bar{I}-\frac{1}{2}%
\beta \mu $ and 
\begin{equation}
Z_{\sigma }=\sum_{\{t,\sigma \}}\exp \left[ \beta
\sum_{j=1}^{N}\sum_{k=1}^{M}t_{j,k}\left( \tilde{\xi}_{k}+J_{k}\sigma
_{j}\sigma _{j+1}\right) +\frac{\bar{I}}{N}\sum_{i,j=1}^{N}\sigma _{i}\sigma
_{j}+\sum_{j=1}^{N}\tilde{h}_{j}\sigma _{j}\right]  \label{zising}
\end{equation}
is the annealed grand partition function of a random Ising model with 
\begin{equation}
\tilde{\xi}_{k}=\xi _{k}+J_{k}
\end{equation}
and 
\begin{equation}
\tilde{h}_{j}\equiv \bar{h}+2\beta \sum_{k=1}^{M}\left[ q_{k}-t_{j,k}\right]
J_{k}.
\end{equation}
Notice that the distributions of the random fields $\tilde{h}_{j}$ and the
short-interactions $\kappa _{j}$ are correlated. The presence of the random
terms in the effective field $\tilde{h}_{j}$ acting on the spin in the $j$th
site breaks the $h\rightarrow $ $-h$, $\{\sigma \}\rightarrow \{-\sigma \}$
symmetry present in the uniform Ising model (for which $J_{k}\equiv J$ for
all $k$). The effect of this on the behavior of the gas is that, unlike the
pure model, the hole-particle symmetry is also broken in the presence of any
randomness, and so all transitions have different critical temperatures.
This is in contrast to the lattice gas with particles interacting via
softened core uniform potentials \cite{hemmer70,stell72}, where multiple
phase transitions are also present, but whose critical temperatures are not
all different because the symmetry is still present.

The correlation between the random fields and random interactions allows one
to perform the partial trace over the disorder variables, mapping the system
onto a regular Ising model with an effective interaction $K/\beta $ and a
interaction dependent effective field. Rewriting Eq.\ref{zising} as 
\begin{equation}
Z_{\sigma }=\sum_{\{\sigma \}}\hat{Z}(\sigma )\prod_{j=1}^{N}\left\{
\sum_{k=1}^{M}\exp \left[ \beta J_{k}\left( \sigma _{j}\sigma _{j+1}-\sigma
_{j}-\sigma _{j+1}\right) +\beta \tilde{\xi}_{k}\right] \right\} ,
\end{equation}
where 
\begin{equation}
\hat{Z}(\sigma )\equiv \exp \left[ \frac{\bar{I}}{N}\sum_{i,j=1}^{N}\sigma
_{i}\sigma _{j}+\left( \bar{h}+2\beta \sum_{k=1}^{M}q_{k}J_{k}\right)
\sum_{j=1}^{N}\sigma _{j}\right] ,
\end{equation}
and performing a decimation 
\begin{equation}
\prod_{j=1}^{N}\left\{ \sum_{k=1}^{M}\exp \left[ \beta J_{k}\left( \sigma
_{j}\sigma _{j+1}-\sigma _{j}-\sigma _{j+1}\right) +\beta \tilde{\xi}_{k}%
\right] \right\} \equiv Ae^{K\left( \sigma _{j}\sigma _{j+1}-\sigma
_{j}-\sigma _{j+1}\right) },
\end{equation}
with 
\begin{equation}
A^{4}=\left( \sum_{k=1}^{M}\tilde{w}_{k}e^{3\beta J_{k}}\right) \left(
\sum_{k=1}^{M}\tilde{w}_{k}e^{-\beta J_{k}}\right) ^{3},
\end{equation}
\begin{equation}
e^{4K}=\left. \left( \sum_{k=1}^{M}\tilde{w}_{k}e^{3\beta J_{k}}\right)
\right/ \left( \sum_{k=1}^{M}\tilde{w}_{k}e^{-\beta J_{k}}\right)
\end{equation}
and 
\[
\tilde{w}_{k}\equiv \exp \left( \beta \tilde{\xi}_{k}\right) , 
\]
one can express the annealed grand-partition function $Z_{\sigma }$ as 
\begin{equation}
Z_{\sigma }=A^{N}\sum_{\{\sigma \}}\exp \left[ K\sum_{j=1}^{N}\sigma
_{j}\sigma _{j+1}+\frac{\bar{I}}{N}\sum_{i,j=1}^{N}\sigma _{i}\sigma _{j}+%
\tilde{h}\sum_{j=1}^{N}\sigma _{j}\right] ,  \label{ziref}
\end{equation}
where 
\begin{equation}
\tilde{h}\equiv \bar{h}+2\left( \bar{I}\sigma +\beta
\sum_{k=1}^{M}q_{k}J_{k}-K\right) .
\end{equation}
By eliminating $A$ and the fugacities $\tilde{w}_{k}$ the annealed free
energy of the random Ising model (as a function of the magnetization) can be
written as 
\begin{eqnarray}
f_{\sigma } &=&f(K,\sigma )+k_{B}T\sum_{k=1}^{M}q_{k}\ln q_{k}+  \nonumber \\
&&+k_{B}T\sum_{k=1}^{M}q_{k}\ln \frac{4e^{K-\beta J_{k}}}{[3-\epsilon
(K)]e^{2(K-\beta J_{k})}+[1+\epsilon (K)]e^{-2(K-\beta J_{k})}},
\label{graclfi}
\end{eqnarray}
in which once more there is an entropy of mixing term and 
\begin{equation}
f(K,\sigma )\equiv -k_{B}T\ln \left\{ e^{K}\left[ \cosh \tilde{h}+\left(
\sinh ^{2}\tilde{h}+e^{-4K}\right) ^{1/2}\right] \right\} +I\sigma ^{2}
\end{equation}
is the free energy of an Ising model with average magnetization $\sigma $ in
a uniform field and uniform short- ($K/\beta $) and long-range ($I$)
interactions \cite{vieira95}, $K$ being determined from the solution of the
equation 
\begin{equation}
\sum_{k=1}^{M}\frac{q_{k}}{2\coth \left[ 2\left( K-\beta J_{k}\right) \right]
+\left[ 1-\epsilon _{\sigma }(K)\right] }=0.
\end{equation}
The nearest-neighbor spin-spin correlation function is explicitly given by 
\begin{equation}
\epsilon _{\sigma }(K)\equiv \left\langle \sigma _{j}\sigma
_{j+1}\right\rangle _{K}=1-\frac{2e^{-4K}\left( \sinh ^{2}\tilde{h}%
+e^{-4K}\right) ^{-1/2}}{\cosh \tilde{h}+\left( \sinh ^{2}\tilde{h}%
+e^{-4K}\right) ^{1/2}}.
\end{equation}

By inverting the transformation \ref{lgim} the annealed free energy of the
lattice gas can be written as 
\begin{equation}
f_{g}=-\frac{k_{B}T}{N}\ln \mathcal{Z}+\sum_{k=1}^{M}q_{k}\xi _{k}+\rho \mu
=\left( h+\sum_{k=1}^{M}q_{k}J_{k}+I\right) +f_{a}+\rho \mu ,
\end{equation}
from where the pressure is calculated by explicitly excluding the entropy of
mixing term present in $f_{a}$, 
\[
P=\rho \mu -f_{g}+k_{B}T\sum_{k=1}^{M}q_{k}\ln q_{k}\Rightarrow 
\]
\begin{equation}
\Rightarrow P=-\left( h+\sum_{k=1}^{M}q_{k}J_{k}+I\right) -f_{\sigma
}+k_{B}T\sum_{k=1}^{M}q_{k}\ln q_{k},  \label{plg}
\end{equation}
while the density is given by 
\begin{equation}
\rho =\frac{1-\sigma }{2}.  \label{dlg}
\end{equation}

By studying the properties of the Ising model described by the free energy $%
f_{\sigma }$ in Eq.\ref{graclfi} the behavior of the lattice gas can be
determined. The equilibrium state of the magnetic system at fixed
temperature and external field $h$ is that which minimizes $f_{\sigma }$. By
locating those minima and using Eqs.\ref{plg} and \ref{dlg} the equation of
state of the gas is exactly determined. This procedure automatically removes
the unstable portions of the isotherms and is equivalent to the Maxwell
construct.
\medskip
\begin{center}
{\Large Appendix B. Ground state properties.}{\large \medskip}
\end{center}
\label{ground}At zero temperature the Helmholtz free energy $F$ of a system
is identical to its internal energy, as it is evident from the thermodynamic
relation 
\begin{equation}
F=E-TS,
\end{equation}
where $T$ is the temperature and $S$ is the entropy.

The ground state behavior of the system can thus be determined by writing
the internal energy $E$ for the various ranges of the density $\rho $ and
calculating the pressure using the expression 
\begin{equation}
\left. P\right| _{T=0}=\mbox{MC}\left[ \rho \frac{\partial E}{\partial \rho }%
-E\right] ,
\end{equation}
where ``MC'' denotes the Maxwell construct \cite{lebowitz66} of the function
in square brackets, which substitutes horizontal parts (corresponding to
phase transitions) for the unstable portions of the isotherms.

A fundamental feature of annealed models is that impurities are free to
rearrange themselves so as to minimize the free energy. For a lattice gas
with random interactions this means that, as particles are added to the
system, interactions of the same kind have a tendency to aggregate, forming
domains. From this analysis it is possible to find those structures which
minimize the internal energy and determine the ground state behavior of the
model discussed in Sec. \ref{slrint}, which is summarized below. In the
following discussion the volume of the system is regarded as fixed.

\textbf{Case }$J_{B}<0\leq J_{A}$. In this situation, because of the
presence of the attractive interactions $J_{A}$, interaction domains are
formed even at infinitesimal densities. For sufficiently weak long-range
attraction energy $I$, the density can be increased from $\rho =0$ to $\rho
=p$ at zero pressure by adding particles to the $J_{A}$ domains. At
densities between $\rho =p$ and $\rho =\frac{1}{2}(1+p)$ new particles can
be allocated with minimum energy cost at every other site of the $J_{B}$
domains. The pressure is nevertheless raised, because of the smaller volume
available. Finally, for $\rho >\frac{1}{2}(1+p)$ particles can be added only
by allocating them in empty sites between two other particles in the $J_{B}$
domain, raising both internal energy and pressure. For strong long-range
attraction some of the transitions are suppressed, the various possibilities
for the $T=0$ isotherms being as follows. 
\[
\begin{array}{c}
\left\{ 
\begin{array}{c}
\mbox{\textbf{3 transitions:}}\vspace{0.15cm} \\ 
\left\{ 
\begin{array}{l}
0<\rho <p,\quad P=0\equiv P_{1}\vspace{0.15cm} \\ 
p<\rho <\frac{1}{2}(1+p),\quad P=4pJ_{A}-2p(1+p)I\equiv P_{2}\vspace{0.15cm}
\\ 
\frac{1}{2}(1+p)<\rho <1,\quad P=4[pJ_{A}-(1-p)J_{B}]-2(1+p)I\equiv P_{3}
\end{array}
\right. \vspace{0.15cm} \\ 
\mbox{condition: }I<\min \left\{ -\frac{2}{1+p}J_{B},\frac{2}{1+p}%
J_{A}\right\}
\end{array}
\right. \\ 
\\ 
\left\{ 
\begin{array}{c}
\mbox{\textbf{2 transitions:}}\vspace{0.15cm} \\ 
\left\{ 
\begin{array}{l}
0<\rho <\frac{1}{2}(1+p),\quad P=P_{1}\vspace{0.15cm} \\ 
\frac{1}{2}(1+p)<\rho <1,\quad P=P_{3}
\end{array}
\right. \vspace{0.15cm} \\ 
\mbox{conditions: }\left\{ 
\begin{array}{l}
p>\frac{J_{A}+J_{B}}{J_{A}-J_{B}}\equiv p_{d}\quad \mbox{and}\vspace{0.15cm}
\\ 
\frac{2}{1+p}J_{A}<I<2\left[ \frac{p}{1+p}J_{A}-J_{B}\right]
\end{array}
\right.
\end{array}
\right. \\ 
\\ 
\left\{ 
\begin{array}{c}
\mbox{\textbf{2 transitions:}}\vspace{0.15cm} \\ 
\left\{ 
\begin{array}{l}
0<\rho <p,\quad P=P_{1}\vspace{0.15cm} \\ 
p<\rho <1,\quad P=4p(J_{A}-J_{B}-I)=P_{2}^{\prime }
\end{array}
\right. \vspace{0.15cm} \\ 
\mbox{conditions: }\left\{ 
\begin{array}{l}
J_{A}>-J_{B}\quad \mbox{and}\vspace{0.15cm} \\ 
p<p_{d}\quad \mbox{and}\vspace{0.15cm} \\ 
-\frac{2}{1-p}J_{B}<I<J_{A}-J_{B}
\end{array}
\right.
\end{array}
\right. \\ 
\\ 
\left\{ 
\begin{array}{c}
\mbox{\textbf{1 transition:} }\vspace{0.15cm} \\ 
\left\{ 0<\rho <1,\quad P=P_{1}\right. \vspace{0.15cm} \\ 
\mbox{conditions: }\left\{ 
\begin{array}{l}
p<p_{d}\quad \mbox{and}\quad I>J_{A}-J_{B}\vspace{0.15cm} \\ 
\mbox{or}\vspace{0.15cm} \\ 
p>p_{d}\quad \mbox{and}\quad I>2\left[ \frac{p}{1+p}J_{A}-J_{B}\right]
\end{array}
\right.
\end{array}
\right.
\end{array}
\]

\textbf{Case }$J_{B}<J_{A}<0\mathbf{.}$ This is the case where all
nearest-neighbor interactions are repulsive. For sufficiently weak
long-range interactions $I$ there is no aggregation of particles at low
densities. On the contrary, particles can be added to the system at minimum
cost by allocating them such that they have no nearest neighbors. In general
this can be done at zero pressure only if $\rho <\frac{1}{2}$. For greater
densities interaction domains are formed, and new particles are allocated,
by raising the pressure, between occupied sites in the $J_{A}$ domains,
which are filled up for $\rho =p+\frac{1}{2}(1-p)=\frac{1}{2}(1+p)$. Adding
further particles requires even higher pressures, since they must occupy
available sites in the $J_{B}\,$domains. Again the existence of the various
possible transitions depends on the strength of the long-range interactions.
Zero temperature isotherms are described below. 
\[
\begin{array}{c}
\left\{ 
\begin{array}{c}
\mbox{\textbf{3 transitions:} }\vspace{0.15cm} \\ 
\left\{ 
\begin{array}{l}
0<\rho <\frac{1}{2},\quad P=0\equiv P_{1}\vspace{0.15cm} \\ 
\frac{1}{2}<\rho <\frac{1}{2}(1+p),\quad P=-4J_{A}-(1+p)I\equiv P_{2}%
\vspace{0.15cm} \\ 
\frac{1}{2}(1+p)<\rho <1,\quad P=4[pJ_{A}-(1-p)J_{B}]-2(1+p)I\equiv P_{3}
\end{array}
\right. \vspace{0.15cm} \\ 
\mbox{condition: }I<\min \left\{ -\frac{4}{1+p}J_{A},4(J_{A}-J_{B})\right\}
\end{array}
\right. \\ 
\\ 
\left\{ 
\begin{array}{c}
\mbox{\textbf{2 transitions:} }\vspace{0.15cm} \\ 
\left\{ 
\begin{array}{l}
0<\rho <\frac{1}{2}(1+p),\quad P=P_{1}\vspace{0.15cm} \\ 
\frac{1}{2}(1+p)<\rho <1,\quad P=P_{3}
\end{array}
\right. \vspace{0.15cm} \\ 
\mbox{conditions: }\left\{ 
\begin{array}{l}
J_{A}>\frac{2}{3}J_{B}\quad \mbox{and}\vspace{0.15cm} \\ 
p>\frac{2J_{A}-J_{B}}{J_{A}-J_{B}}\equiv p_{s}\quad \mbox{and}\vspace{0.15cm}
\\ 
-\frac{4}{1+p}J_{A}<I<2\left[ \frac{p}{1+p}J_{A}-J_{B}\right]
\end{array}
\right.
\end{array}
\right. \\ 
\\ 
\left\{ 
\begin{array}{c}
\mbox{\textbf{2 transitions:} }\vspace{0.15cm} \\ 
\left\{ 
\begin{array}{l}
0<\rho <\frac{1}{2},\quad P=P_{1}\vspace{0.15cm} \\ 
\frac{1}{2}<\rho <1,\quad P=-4[pJ_{A}+(1-p)J_{B}+I]=P_{2}^{\prime }
\end{array}
\right. \vspace{0.15cm} \\ 
\mbox{conditions: }\left\{ 
\begin{array}{l}
J_{A}<\frac{1}{2}J_{B}\quad \mbox{and}\vspace{0.15cm} \\ 
p<p_{s}\quad \mbox{and}\vspace{0.15cm} \\ 
4(J_{A}-J_{B})<I<-2[pJ_{A}+(1-p)J_{B}]
\end{array}
\right.
\end{array}
\right. \\ 
\\ 
\left\{ 
\begin{array}{c}
\mbox{\textbf{1 transition:} }\vspace{0.15cm} \\ 
\left\{ 0<\rho <1,\quad P=P_{1}\right. \vspace{0.15cm} \\ 
\mbox{conditions: }\left\{ 
\begin{array}{l}
p<p_{s}\quad \mbox{and}\quad I>-2[pJ_{A}+(1-p)J_{B}]\vspace{0.15cm} \\ 
\mbox{or}\vspace{0.15cm} \\ 
p>p_{s}\quad \mbox{and}\quad I>2\left[ \frac{p}{1+p}J_{A}-J_{B}\right]
\end{array}
\right.
\end{array}
\right.
\end{array}
\]

\newpage

{\LARGE Figure captions\medskip }

\noindent \textbf{Figure 1.} (a-c) $\theta \times \alpha $ diagram, where $%
\theta \equiv k_{B}T/\left| J_{B}\right| $ is the renormalized temperature
and $\alpha \equiv I/\left| J_{B}\right| $, for dilute short-range
interactions ($J_{A}=0$) with probabities $p=0.1$, $p=0.25$ and $p=0.5$,
respectively. Continuous curves represent critical temperatures ($\theta
_{c} $ and $\theta _{c}^{\prime }$) of the phase transitions, while triple
point temperatures ($\theta _{t}$) are shown as dotted curves. Curves for $%
\theta _{c}^{\prime }$ and $\theta _{t}$ coincide at point $P_{ct}$, whose $%
\alpha $ coordinate ($\alpha _{ct}$) is shown in (d) as a function of $p$.%
{\LARGE \medskip }

\noindent \textbf{Figure 2. }(a) Isotherms $P^{*}\times \rho $ (continuous
curves), where $P^{*}\equiv P/\left| J_{B}\right| $ is the renormalized
pressure, and coexistence curves (dotted curves) for dilute short-range
interactions ($J_{A}=0$), $p=0.1\,$and $\alpha =1.5$; (b) Phase diagram $%
P^{*}\times \theta $ showing close-packed (CP), liquid (L) and gas (G)
phases. Continuous curves correspond to first-order transitions terminating
at the critical temperatures $\theta _{c}$ and $\theta _{c}^{\prime }$. 
{\LARGE \medskip }

\noindent \textbf{Figure 3. }(a) The same as in Fig. 2(a) for $\alpha =2.2$;
(b) The same as in Fig. 2(b), but with a triple point indicating coexistence
of the three phases.{\LARGE \medskip }

\noindent \textbf{Figure 4. }(a) The same as in Fig. 2(a) for $\alpha
_{ct}\simeq 2.4297$; (b) Phase diagram showing the coincidence of the triple
point (temperature $\theta _{t}$) and the critical point of the
liquid-close-packed (temperature $\theta _{c}^{\prime }$). {\LARGE \medskip }

\noindent \textbf{Figure 5. }$\rho \times \theta $ projections of the
coexistence surface for $\delta =1$ ($\delta \equiv J_{A}/\left|
J_{B}\right| $), $p=0.07$ and (a) $\alpha =1.13$ and (b) $\alpha =1.129$.
The triple point present in (a), whose temperature is $\theta _{t}$, is
suppressed in (b), where a coexistence ``bubble'' can be observed, as
indicated by the arrows.{\LARGE \medskip }

\noindent \textbf{Figure 6. }$\theta \times \alpha $ diagram for $\delta =1$
and $p=0.07$. The inset shows the breaking and reentrant behavior of the $%
\theta _{c}$ curve, as well as the existence of the new points $%
P_{ct}^{\prime \prime }$ and $P_{ct}^{\prime \prime \prime }$, extremes of a
new triple point line (dotted). The $\theta _{c}^{\prime }$ curve is ommited
from the inset for clarity.{\LARGE \medskip }

\noindent \textbf{Figure 7. }$\theta \times \alpha $ diagram for $\delta =2$
and $p=0.1$. The four (dotted) triple point lines intersect at the quadruple
point $P_{q}$.{\LARGE \medskip }

\noindent \textbf{Figure 8. }Phase diagram for $\delta =2$, $p=0.08$ and (a) 
$\alpha =2.3$, (b) $\alpha _{q}\simeq 2.3277$ and (c) $\alpha =2.35$. Two
triple points are evident in (a) and (c), while the quadruple point is
present in (b). In all cases the axes intervals are $0\le \theta \le 2.4$
and $0\le P^{*}\le 0.8$. {\LARGE \medskip }

\noindent \textbf{Figure 9. }$\rho \times \theta $ ($\theta \equiv
k_{B}T/\left| J_{B}\right| $) projections of the coexistence surfaces for $%
\delta =2$, $p=0.08$ and (a) $\alpha =2.3$, (b) $\alpha _{q}\simeq 2.3277$
and (c) $\alpha =2.35$. The broken lines indicate the triple point
temperatures in (a) and (c), and the quadruple point temperature in (b).%
{\LARGE \medskip }

\noindent \textbf{Figure 10. }(a-c) $\theta \times \alpha $ diagram, where $%
\theta \equiv k_{B}T/\left| J_{B}\right| $ is the renormalized temperature
and $\alpha \equiv I/\left| J_{B}\right| $, for the case $J_{A}=\frac{3}{10}%
J_{B}<0$ ($\delta =-\frac{3}{10}$) with probabities $p=0.25$, $p=0.25$ and $%
p=0.5$, respectively. Continuous curves represent critical temperatures ($%
\theta _{c}$ and $\theta _{c}^{\prime }$) of the phase transitions, while
triple point temperatures ($\theta _{t}$) are shown as dotted curves.
Coordinates $\alpha _{ct}$ and $\alpha _{ct}^{\prime }$ of points $P_{ct}$
and $P_{ct}^{\prime }$, as well as values $\alpha _{2}$ and $\alpha _{3}$ at
which triple point temperatures become zero, are shown in (d) as a function
of $p$. Notice that, while $\alpha _{ct}$ goes to $\alpha _{tr}\simeq 3.1532$
as $p\rightarrow 0$, $\alpha _{ct}^{\prime }$ goes to $\left| \delta \right|
\alpha _{tr}\simeq 0.9459$ as $p\rightarrow 1$, as expected.{\LARGE \medskip 
}

\noindent \textbf{Figure 11. }(a) $P^{*}\times \rho $ isotherms (solid
lines) and coexistence curves (dotted lines) for $\delta =-0.3$, $p=0.5$ and 
$\alpha =0.8$; (b) $P^{*}\times \theta $ phase diagrams showing the various
transition lines and critical temperatures; (c) $\rho \times \theta $
projections of the coexistence curves shown in (a).{\LARGE \medskip }

\noindent \textbf{Figure 12. }The same as in Fig. 11 for $p=0.75$. Notice in
(b) and (c) the existence of a triple point.{\LARGE \medskip }

\noindent \textbf{Figure 13. }$\theta \times \alpha $ diagram for $\delta
=-0.6$ and $p=p^{\dagger }=0.5$, showing the critical temperatures (solid
lines) and triple point temperatures (dotted lines).{\LARGE \medskip }

\noindent \textbf{Figure 14. }(a) $P^{*}\times \rho $ isotherms (solid
lines) and coexistence curves (dotted lines) for $\delta =-0.6$, $p=0.5$ and 
$\alpha =1.45$; (b) $P^{*}\times \theta $ phase diagrams showing the various
transition lines, critical temperatures and the triple point.

\end{document}